\documentclass[aps,twocolumn,prl,amsmath,amssymb]{revtex4-1}

\usepackage{graphicx,epsfig}
\usepackage{color}
\usepackage{wasysym}
\usepackage[normalem]{ulem}
\usepackage{bm}

\begin{document}
\title{Periodically Driven Many-Body Systems:\\ A Floquet Density Matrix Renormalization Group Study}
\author{Shaon Sahoo$^{1,2}$}
\author{Imke Schneider$^1$}
\author{Sebastian Eggert$^1$}
\affiliation{$^1$Physics Department and Research Center OPTIMAS, University of Kaiserslautern, 67663 Kaiserslautern, Germany}
\affiliation{$^2${Department of Physics, Indian Institute of Technology Tirupati, Tirupati 517506, India}}

\parindent = 0pt

\begin{abstract}
Driving a quantum system periodically in time can profoundly
alter its long-time correlations and give rise to exotic quantum states of matter. 
The complexity of the combination of 
many-body correlations and dynamic manipulations has the potential to 
uncover a whole field of new phenomena, but the theoretical and numerical understanding 
becomes extremely difficult.  We now 
propose a promising numerical method by generalizing the density matrix renormalization group
to a superposition of Fourier components of periodically driven many-body systems
using Floquet theory.
With this method we can 
study the full time-dependent quantum solution in a large parameter range for all evolution
times,  beyond the commonly used 
high-frequency approximations.  
Numerical results are presented for the
isotropic Heisenberg antiferromagnetic spin-1/2 chain under both local
(edge) and global driving for spin-spin correlations and temporal
fluctuations. As the frequency is lowered, we demonstrate that more and more Fourier 
components become relevant and determine 
strong length- and frequency-dependent changes of the quantum correlations that cannot be
described by effective static models.
\end{abstract}

\maketitle

{\it Introduction} -- 
Time-periodically driven quantum systems exhibit many interesting macroscopic phenomena, like 
dynamical localization \cite{kaya08,nag14}, 
coherent destruction of tunneling \cite{grifoni98}, 
topological edge modes \cite{cherpakova},
dynamical phase transitions \cite{prosen11,shirai14}, 
quantum resonance catastrophe \cite{thuberg16,reyes17}, 
and time crystals \cite{else16,jzhang17,choi17,kreil}.
Driven systems can hold new non-equilibrium phases which do not have counterparts in the 
equilibrium \cite{khemani16}. From a practical point of view, periodic driving can be used to control the magnetic 
order \cite{gorg18,wang}, create new quantum topological states and Majorana end modes \cite{rudner13,goldman14,thakurathi13}, 
construct a perfect spin filter \cite{thuberg17},
or obtain transient superconducting behavior well above the transition temperature \cite{mitrano16}. 

From a theoretical point of view, time-periodic quantum systems can conveniently be described
using Floquet theory \cite{grifoni98}, which basically analyzes the coupled
Fourier components using an additional discrete Floquet dimension. 
Analytically, this Floquet 
approach allows 
a systematic high-frequency expansion and an effective quasi-static description \cite{eckardt15, itin15,wang14,rahav03}.
However, the numerical density matrix renormalization group (DMRG) method, which is 
highly successful for studying one dimensional correlated
systems in equilibrium \cite{white92,scholl05}, is yet to 
find a satisfactory extension in this regard.
Using time-dependent DMRG a real-time approach was given in Refs.~\onlinecite{poletti11, kennes18}. 
An analysis of the 
high frequency range  becomes possible in this way 
albeit with growing error for longer evolution times. 
This problem can be avoided by considering only
stroboscopic time-evolution, since
the 
time evolution operator of period $T$ can be expressed using matrix product states \cite{czhang17}.
Despite these advances,  
a full solution for all times
of the steady state of a periodically driven many-body model using DMRG
is still elusive.  We now propose to use the DMRG method directly in Floquet space, 
which consists of infinitely many copies of the original many-body Hilbert space.
We therefore have to solve the problem of finding a reasonable truncation in 
Floquet index and develop a targeting algorithm, which takes into account that
there is no ground state for the unbounded Floquet quasi-spectrum.
However, by overcoming these obstacles the payoff is rewarding:  The resulting eigenstates
are time-periodic steady states, so emerging many-body 
correlations can 
be studied on any time scale, including in the infinite time limit or 
time averaged.  The main advantage of DMRG to accurately 
capture entanglement along
the system is fully maintained despite the extension to an 
additional Floquet dimension.
For demonstration we study the prototypical model of the 
antiferromagnetic Heisenberg spin chain,  which is subject to 
either local (edge) or global time periodic driving.  As a function of frequency we
observe strong size-dependent changes of correlations, 
which are not captured by the analytically
derived high-frequency effective model.

{\it Theoretical background:} -- 
Let us first present the general theoretical background of the 
Floquet eigenvalue equations for a time-periodic 
Hamiltonian, 
\begin{eqnarray}\label{tdham}
H(t)=H_0+2\mu H_1 \textrm{cos} (\omega t),
\end{eqnarray}
where $H_0$ is the time-independent Hamiltonian and $H_1$ 
represents the coupling to an external time-periodic driving
of frequency $\omega$ and strength $\mu$, which is chosen monochromatic, 
but can in principle be generalized to any time-periodic Hamiltonian $H(t)=H(t+T)$ with
 $T=2\pi/\omega$. In this case,
steady-state solutions of the Schr\"odinger equation can be written in the form
$|\Psi(t)\rangle = e^{-i\epsilon t}|\Phi(t)\rangle$, where $\epsilon$ is the Floquet 
quasi-energy and 
$|\Phi(t)\rangle=|\Phi(t+T)\rangle$ 
is the time-periodic Floquet mode \cite{sambe73}, 
which 
can be decomposed in a Fourier series, 
\begin{eqnarray}\label{frdecom}
|\Phi(t)\rangle=\sum_{n=-\infty}^{\infty}e^{-in\omega t} |\Phi_n \rangle.
\end{eqnarray}
 The time-dependent Schr\"odinger equation then becomes an eigenvalue equation
$(H(t)-i\partial_t)|\Phi(t)\rangle=\epsilon |\Phi(t)\rangle$ \cite{sambe73}, corresponding
to  a set of coupled equations
$(H_0 -n\omega)|\Phi_n \rangle + \mu H_1 (|\Phi_{n-1}\rangle + |\Phi_{n+1} \rangle)=\epsilon |\Phi_n \rangle$, 
in terms of the Fourier components.
Hence we need to solve an infinite dimensional eigenvalue equation
${\cal H} C = \epsilon C$, where $\cal H$, the Floquet 
matrix, is a tridiagonal block matrix,
\begin{eqnarray}
  {\cal H} = 
  \begin{bmatrix}
     ~~~\ddots & ~ & ~ & ~&~&  \\
    \mu H_1 & H_0-\omega\bm{1}& \mu H_1 & ~ &~& \\
    ~ & \mu H_1 & H_0 & \mu H_1 & ~&\\
    ~ & ~ & \mu H_1 & H_0+\omega\bm{1} & ~\mu H_1 &\\
    ~ & ~ & ~ & ~ & \ddots &  
  \end{bmatrix}.
  \label{blktdmt}
\end{eqnarray} 
Here each block has the dimension of the many-body Hilbert space and
$C= (\cdots |\Phi_{1} \rangle |\Phi_{0} \rangle |\Phi_{-1} \rangle \cdots)^T$ is 
the block column vector representing all Fourier components sequentially, 
which is normalized $CC^\dagger= \sum_{n=-\infty}^{\infty} \langle \Phi_n|\Phi_n \rangle= 1$ \cite{hanggi88}.
Note, that a trivial change of Floquet index by an integer $n\to n+\ell$ simply leads to a 
shift of the quasi-energy $\epsilon \to \epsilon+\ell \omega$, but an otherwise equivalent
solution.  Hence, $\epsilon$ is only defined modulo $\omega$, analogous to a Brillioin zone
in Floquet space. Unfortunately, this also implies that the eigenvalue
$\epsilon$ cannot be assumed to be bounded
above or below. 

\begin{figure}[t]
\includegraphics[angle=270,width=\columnwidth]{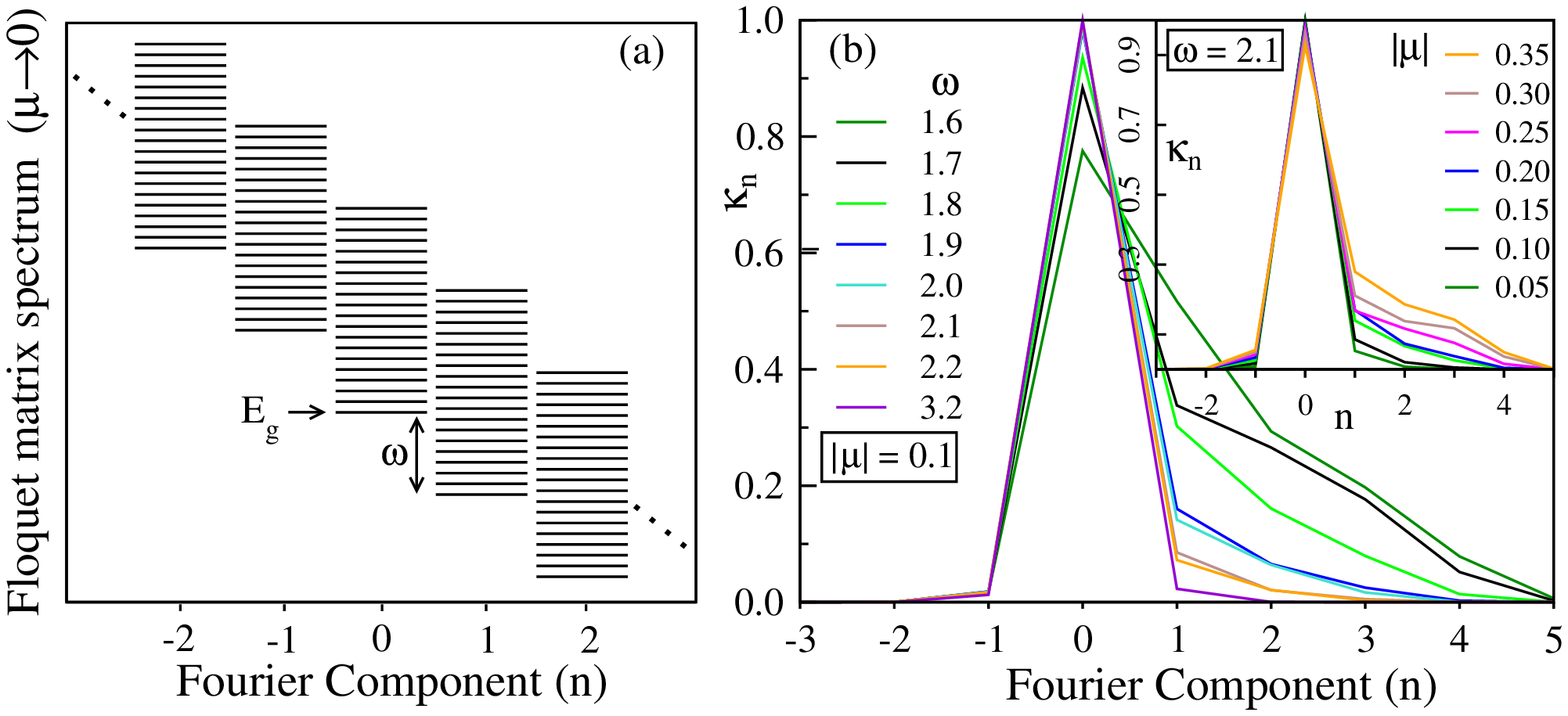}
\caption{(a) Schematic spectrum of $\cal H$ for independent blocks $\mu\rightarrow0$,
 where $E_g$ is the groundstate energy of $H_0$. 
(b) Norms of Fourier components ($\kappa_n = \Vert |\Phi_n \rangle \Vert$) 
from Floquet DMRG simulations of Eq.~(\ref{hs_ld}) 
 for different $\omega$ and different $\mu$ (inset) in units of $J$ 
using $N=30$. }
\label{fig1}
\end{figure}

Even though our method does {\it not} rely on a perturbative approach, it is instructive to 
consider the case of vanishing external driving $\mu\to 0$, which 
leads to a simple shift of an identical spectrum 
for each component as depicted in Fig.~\ref{fig1}.
In this static case, 
$|\Phi(t)\rangle = |\Phi_0 \rangle$ 
will be time independent 
with all other components 
$|\Phi_{n\ne 0} \rangle=0$. 
Staring e.g. from the ground state with eigenenergy $\epsilon =E_g$, increasing the amplitude $\mu$ will lead to an occupation
of more and more components around $n=0$ at a given $\omega$.
On the other hand, lowering $\omega$ will involve more and more excited states with
eigenenergy $\epsilon'$, which become degenerate in the Floquet energy 
$\epsilon = \epsilon'-n \omega$ as depicted in Fig.~\ref{fig1}a. 
This illustrates that for smaller $\omega$ and larger $\mu$ it is important to keep 
more and more Fourier components in the numerical simulations as illustrated in
Fig.~\ref{fig1}b.

{\it Truncation of $\cal H$} --
There are now two types of truncation required:  The DMRG truncation of the 
original Hilbert space in each block 
as well as a truncation in the number of Fourier components
in Eq.~(\ref{blktdmt}).
Note, however, that there is no entanglement between different 
Fourier components
along the Floquet dimension, so the truncation in the number of Fourier components is
independent of the DMRG procedure and
solely based on the choice of amplitude $\mu$ and frequency $\omega$.
We find that the most efficient approach is to retain a fixed number 
of Fourier components between $n=-M_1$ to $M_2$ as numerically feasible.
The results are considered reliable 
as long as we keep all important components, which have significant weight
$\Vert |\Phi_n \rangle \Vert/\Vert |\Phi_0 \rangle \Vert>c$, where $c$ 
is an accuracy factor which limits the truncation error to be of order $c^2$.  
Since typically $\Vert |\Phi_{\pm |n|} \rangle \Vert \ge \Vert 
|\Phi_{\pm (|n|+1)} \rangle \Vert$ it is easy to see when the highest components 
become too large, which accordingly limits the range of parameters.
For lower values of $\omega$ it is useful to choose $M_2 > M_1$, as is evident from Fig. \ref{fig1}b.
In our case, we have taken moderate values of 
$M_1 = 3$ and  $M_2 = 5$ for a total of nine components 
which give an accuracy of $c<10^{-4}$ for the 
results presented below.

{\it Floquet DMRG } -- 
As in the original DMRG algorithm \cite{white92,scholl05}, in the first step 
a small system size is considered, 
so a desired eigenstate can be obtained exactly.  In the static
case the ground state can be found by standard Lanczos {or Davidson} methods, but for the Floquet matrix
$\cal H$ in Eq.~(\ref{blktdmt}) this would result in a lowest quasi-energy state 
which is dominated by the largest possible
{Fourier component} $M_2$ in Fig.~\ref{fig1}a, which is not at all what we want {since this 
situation corresponds to large truncation error}.
Instead the target state must be adiabatically connected to the ground state, 
which has the strongest weight in the $n=0$ sector in
the middle of the quasi-energy spectrum in Fig.~\ref{fig1}a.
We therefore adopt the {\it shift-and-square} method with a dynamical
shift.
For diagonalization of the shifted and squared matrix $({\cal H}-\sigma\bm{1})^2$, we use Davidson's algorithm \cite{davidson75} suitably modified for 
the present problem. 
The matrix $({\cal H}-\sigma\bm{1})^2$ is first projected onto a small search space,
where the shift $\sigma$ takes the groundstate energy $E_g$ as its initial value.
The small projected matrix is then fully solved and its eigenstate with largest overlap with the groundstate is picked 
up. This gives the first approximation of the desired eigenstate  of $\cal H$, which 
in turn is used, by the correction vector method prescribed by Davidson, to enlarge
the search space and then the whole process is repeated for this larger space. 
If the search space becomes inconveniently large, we restart the iterative process 
with the latest solution and replace the shift $\sigma$ by the approximate quasi-energy found at 
this stage (see Appendix for more details). Although $\cal H$
in Eq.~(\ref{blktdmt}) is a big matrix,
it is possible
to simplify the calculation by only manipulating smaller matrices $H_0$ and $H_1$.

We are now in the position to apply the DMRG algorithm, which is based on defining a 
suitable reduced density matrix from the previously targeted eigenstate, so that 
the basis set can be projected to the most important states before the system is
enlarged \cite{white92,scholl05}.  Since the steady state solution $|\Psi(t)\rangle$, 
all Fourier components $|\Phi_n\rangle$
as well as the static ground state are described by the same Hilbert space, we seek a suitable DMRG procedure, 
which projects all blocks in Eq.~(\ref{blktdmt}) simultaneously 
by the same transformation.
Therefore the DMRG projection is found by truncating the original Hilbert space in the same
way as in the static case, albeit using a different Floquet target state and 
corresponding reduced density matrix.

To illustrate the choice of density matrix for a given Floquet eigenstate
$|\Psi (t) \rangle = e^{-i\epsilon t} \sum_{n=-\infty}^{\infty}e^{-in\omega t} 
|\Phi_n \rangle$, we write the Hilbert 
space as the product of two parts -- system ``S" and environment ``E" -- and consider the 
calculation of the time average 
of a local operator $A_S\otimes I_E$ acting on the system block  ``S", where $I_E$ is
the identity
operator for the environment block ``E"
\begin{equation}
\bar{A}_S=\frac{1}{T}\int_0^T \langle \Psi (t)| A_S \otimes I_E | \Psi (t) \rangle dt =
Tr \{\rho_F A_S\}. 
\end{equation}
Here $\rho_F=\sum_{n=-\infty}^{\infty} \theta_n \rho_n$ and $\rho_n$ is the reduced density matrix of the block ``S" for the normalized 
Fourier component $\frac{1}{\sqrt{\theta_n}}|\Phi_n \rangle$ with $\theta_n = \langle \Phi_n |\Phi_n \rangle$.
Due to the normalization 
of a Floquet mode, $ \sum_{n=-\infty}^{\infty} \theta_n =1$, $\rho_F$ will have properties of a density matrix. 
In particular, if
$ \rho_F |r_i \rangle = r_i |r_i \rangle$, 
then $r_i \ge 0$ and $\sum_{i=1}^{D_S} r_i = 1$ where $D_S$ is the dimension of the state
space of ``S". 

Since we need the information of the groundstate 
of $H_0$ to find the Floquet target state, it is
necessary to also consider the reduced density matrix of the groundstate $\rho_G$
in a linear combination 
$\rho=\lambda\rho_F+(1-\lambda)\rho_G$, which is then used in the renormalization.
Here $0<\lambda<1$ and its optimal value can depend on the parameter regime we are 
working in. 
{For low values of $\omega$ and large $|\mu|$, $\lambda$ plays a 
{significant} role.  We find that the value of $\lambda=3/4$ gives a 
good balance
between $\rho_F$ and $\rho_G$ in this regime.
For larger frequencies the results become
insensitive to the value of $\lambda$ since $\rho_F$ and $\rho_G$ are very close, so 
we keep $\lambda=3/4$ in all calculations.}

In summary we have achieved the following Floquet DMRG procedure
for the steady state eigenvalue problem:
(i) For a small system 
(e.g.~6 sites), form $H_0$, $H_1$, and $\cal H$ in Eqs.~(\ref{tdham}) and (\ref{blktdmt})
keeping finite number of Fourier modes ($M_2>M_1$).
(ii) Find the static groundstate vector and energy of $H_0$.
(iii)  Find the Floquet mode with largest overlap with the
groundstate, i.e.~the eigenstate of $\cal H$ following the {\it shift-and-square} method 
described above. 
(iv) Divide the full system into two blocks -- ``S" and ``E" (full system = $S E$). 
Form the reduced density matrix 
$\rho=\lambda\rho_F+(1-\lambda)\rho_G$ for ``S" from the Floquet mode and the ground state. 
Diagonalize $\rho$ and retain $M$ most significant eigenvectors to project 
the relevant operators of ``S" and ``E" into this subspace. 
(v) Enlarge the system by adding two sites between the two blocks 
(new full system = $S\bullet\bullet E$). 
Form $H_0$ and $H_1$ for the extended system (``superblock") to construct the 
corresponding $\cal H$. 
(vi) Go to the step (ii) and repeat until the system size reaches the target size.
{In all our simulations we use $M_1 = 3$, $M_2 = 5$, and $M=180$, 
giving a total 
superblock dimension of $4M^2(M_1+M_2+1)\agt 10^6$ for the Floquet matrix $\cal H$, 
which is somewhat smaller 
than in ordinary DMRG calculations due to the more involved search algorithm for the
target state in the middle of the spectrum.}


{\it Results} --
Using the Floquet DMRG method we study both locally (edge) and globally driven spin-1/2 Heisenberg antiferromagnetic chains (isotropic) 
with $N$ spins
\begin{align} 
H_{ld}(t)&=J\sum_{i=1}^{N-1}\vec{S}_i\cdot\vec{S}_{i+1} +2\mu~\textrm{cos} (\omega t) S_{1}^{z} ~~\textrm{and} \label{hs_ld}\\
H_{gd}(t)&=J\sum_{i=1}^{N-1}\vec{S}_i\cdot\vec{S}_{i+1} + 2\sum_{j=1}^{N}\mu_j ~\textrm{cos} (\omega t) S_{j}^{z} \label{hs_gd}. 
\end{align}
Here $\vec{S}_i$ is the $i^\text{th}$ spin and $S_{j}^{z}$ is the $z$-component of the $j^\text{th}$ spin. 
In Eq.~(\ref{hs_gd}) an incommensurate  modulation is applied with
$\mu_j=\mu\,\textrm{cos}~2\pi\beta j$ for $j\le N/2$ and $\mu_j=\mu \,\textrm{cos}~2\pi\beta (N+1-j)$ for $j> N/2$ using
$\beta=211/311$ in order to produce a 
dynamical pseudo-disorder for the globally driven system. 

For comparison an effective high frequency model can be derived based on the exact solution 
of the corresponding Ising model without spin-flip with
Ising eigenstates $|\{{s_j^z}\}\rangle=\prod_{\otimes j} |{s_j^z}\rangle$ where   $|s_j^z\rangle$ represent  the local $s_j^z$-basis states.  The Floquet modes can then be exactly determined to be (see Appendix)
\begin{align}\label{floquet_modes_ising}
|\Phi(t)\rangle= \prod_{\otimes j} e^{- i \frac{2 \mu_j}{ \omega}s_j^z \sin(\omega t)} |s_j^z\rangle
\end{align} with quasi-energies 
$\epsilon(\{s_j^z\})=J\sum_j s_j^z s_{j+1}^z$
in terms of $s_j^z$ quantum numbers.
The Fourier decomposition in Eq.~(\ref{frdecom})   yields
$|\Phi_n\rangle=J_{-n}\left(\frac{2}{ \omega}\sum_j \mu_j s_j^z\right) \prod_{\otimes j}  |s_j^z\rangle,$
where $J_{n}(x)$ denote Bessel functions of the first kind. 
An effective static Hamiltonian is then derived in 
a high frequency approximation $\omega\gg J$
\cite{eckardt15, itin15,wang14,rahav03} by perturbatively calculating
the time-averaged matrix elements of the full Floquet Hamiltonian $(H(t)-i\partial_t)$ 
with respect to the Ising Floquet modes in Eq.~(\ref{floquet_modes_ising}) and
taking into account only terms in the $n=0$ sector  (see Appendix)
\begin{equation}\label{h_effective} 
	H_{\rm eff}=\sum_{j=1}^{N-1} \left[{J_{j,j+1}}\!\left(S^x_j S^x_{j+1} \!+\!S^y_j S^y_{j+1}\right)+ J S^z_jS^z_{j+1}\right]  
	\end{equation}
	where $J_{j,j+1}=J J_0\left(\frac{2 (\mu_j-\mu_{j+1})}{\omega}\right)$. 
The comparison with this effective model allows a systematic analysis of the effect of
higher Fourier modes on correlations in DMRG as the frequency is lowered.

\begin{figure}[t]
\includegraphics[height=\columnwidth,angle=270]{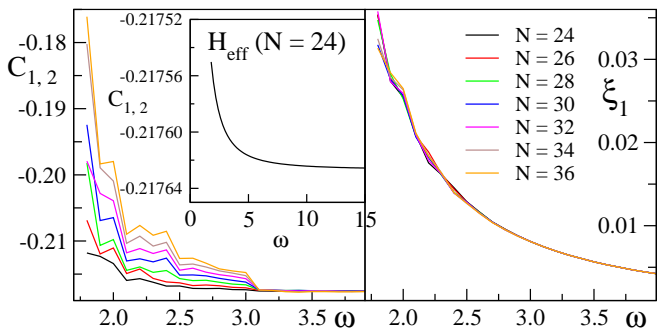}
\caption{{Floquet DMRG} results for the correlations and fluctuations of the 
edge driven Heisenberg model in Eq.~(\ref{hs_ld}) for $\mu=0.1J$. The correlation result
for the static effective Hamiltonian.}
\label{fig2}
\end{figure}
We first consider the locally driven model in Eq.~(\ref{hs_ld}) using the 
Floquet DMRG.
Lowering the 
frequency from $\omega/J = 3.2$ to $1.6$ we observe a significant occupation
in higher Fourier components in Fig.~\ref{fig1}b.  This signals a rather sudden
crossover from a high frequency regime described by Eq.~(\ref{h_effective}) to
a Floquet regime. This is also reflected in the spin correlations 
\begin{equation}
C_{i,j}=\frac{1}{T}\int_0^T\langle \Phi(t)|S_{i}^zS_{j}^z|\Phi(t)\rangle dt
\end{equation}
As shown in Fig.~\ref{fig2} ({inset}) for $C_{1,2}$ the effective model in Eq.~(\ref{h_effective})
predicts a
reduction of only 0.03\% over this frequency range for $\mu=0.1J$
while the DMRG shows already a 100 times larger
change for $N=24$. Moreover, the edge correlation $C_{1,2}$ shows a surprisingly 
strong dependence on site number $N$, which is well beyond conventional 
renormalization scenarios \cite{affleck,rommer}.  
The unexpected length dependence is caused by the reduced level spacing with increasing $N$,
which facilitates a coupling and hybridization 
with an exponentially increasing number of higher energy states \cite{schneider}
for this parameter range.  The spin fluctuation
$\xi_1=\sqrt{[}\frac{1}{T}\int_0^T(\langle s_{1}^z(t)\rangle -\overline{s}_{1}^z)^2dt]$
also increases quickly with lowered
frequency as shown in Fig.~\ref{fig2}, but does not show the same dramatic dependence on $N$.
The accuracy of the DMRG, the dependence on states kept $M$, the
overlap with the ground state, as well as the behavior of the quasi-energy $\epsilon-E_g$
are discussed in the Appendix.
 
\begin{figure}[t]
\includegraphics[angle=270,width=\columnwidth]{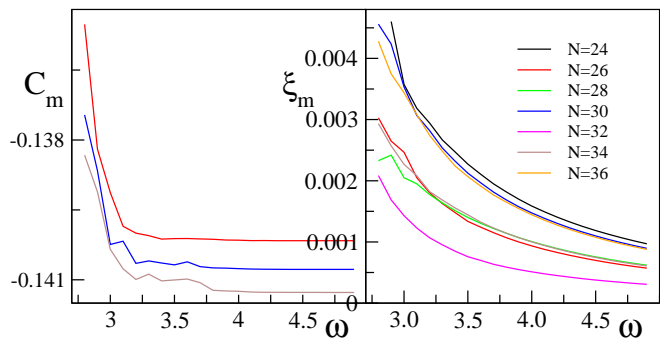}
\caption{{Floquet DMRG} results for the correlations and fluctuations in the middle (m) of the 
bulk pseudo-randomly driven Heisenberg model in Eq.~(\ref{hs_gd}) for $\mu=0.02J$.}
\label{fig3}
\end{figure}

We now turn to the pseudo-randomly driven system in Eq.~(\ref{hs_gd}) for $\mu=0.02J$.
In this case,
we consider the average correlations in the middle (m) of the chain 
${C}_{\rm m} = \frac13 \sum_{i=N/2-1}^{N/2+1}C_{i,i+1}$
and their corresponding fluctuations
${\xi}_{\rm m} = \sqrt{[}\frac14  \sum_{i=N/2-1}^{N/2+2}\xi_i^2]$ in Fig.~\ref{fig3}.
Again we observe a very quick change below frequencies $\omega\alt 3J$, 
but the length dependence is much less dramatic in this case.  Note, that the 
dependence on $N$ for high frequencies can be attributed to the slight shift 
in local values of $\mu_j$ due to the incommensurate modulation.


{\it Conclusion} -- 
We have shown that time-periodically driven many-body systems can be treated 
by a specifically adapted Floquet DMRG method.  The main technical 
difficulty is the targeting of a Floquet mode, 
which is adiabatically connected to the ground state, but 
has a quasi-energy in the middle of the 
Floquet spectrum. 
This problem can be tackled by using a {\it shift-and-square} method
in combination with Davidsons algorithm.  Overall this limits 
the number of states $M$ which can be kept in the Floquet DMRG procedure
compared to static ground state problems, but the accuracy is still very good 
over a wide parameter range,
which allows to determine the emergent 
many-body correlations directly in the infinite-time steady-state limit.

It must be emphasized that it is {\it a priori} unclear which many-body systems 
will show the most interesting Floquet-induced correlations or dynamic 
phase transitions. 
A reliable but straight-forward 
numerical method such as this Floquet DMRG will therefore help to identify 
and classify promising correlated models.  To initiate the search we have chosen the most 
obvious and maybe oldest \cite{bethe} prototypical model of a Heisenberg spin chain.
Time periodic driving is applied at the edge as well as in the bulk with a 
pseudo random distribution.  
For the edge driven system deviations from the effective high-frequency regime
quickly occur starting below $\omega\alt 3J$ as can be seen in the occupation
of higher {Fourier components} in Fig.~\ref{fig1} 
and the drop in edge correlations in Fig.~\ref{fig2} with an unexpectedly strong
dependence on $N$.
The global pseudo-random driving also shows a significant low-frequency 
change starting at approximately the same frequency $\omega\alt 3J$.
Increasing the amplitude $\mu$ will shift this cross-over frequency to slightly 
lower values, but as can be seen in the inset of Fig.~\ref{fig1}b the effect 
of changing $\mu$ is overall less pronounced than a change in $\omega$.

While these results are interesting, they also still lack a better understanding
which may be found by comparison to a broader range of other relevant many-body 
systems in the future.  
We therefore hope that 
the proposed Floquet DMRG will provide a valuable tool to deepen the understanding 
of emergent many-body correlations
from time-periodic driving.

\begin{acknowledgments} -- 
We acknowledge the support from the Deutsche Forschungsgemeinschaft (DFG) 
via the collaborative research centers SFB/TR173 and SFB/TR185. 
\end{acknowledgments} --

\widetext
\section{Appendix} \label{SM}


\setcounter{equation}{0}
\setcounter{figure}{0}
\setcounter{table}{0}
\makeatletter
\renewcommand{\theequation}{S\arabic{equation}}
\renewcommand{\thefigure}{S\arabic{figure}}
\renewcommand{\bibnumfmt}[1]{[S#1]}
This Appendix provides additional data on the quasi-energies and the wave-function overlap,  a detailed discussion of the algorithm to find the eigenstate of 
the Floquet matrix $\cal H$, 
 details on the numerical performance and error as a function of $M$ of the proposed DMRG method, as well as the derivation of the effective Hamiltonian in the large frequency limit.

\subsection{Change of quasi-energies and wave-function overlap}
\begin{figure}[b]
\includegraphics[height=0.49\columnwidth,angle=270]{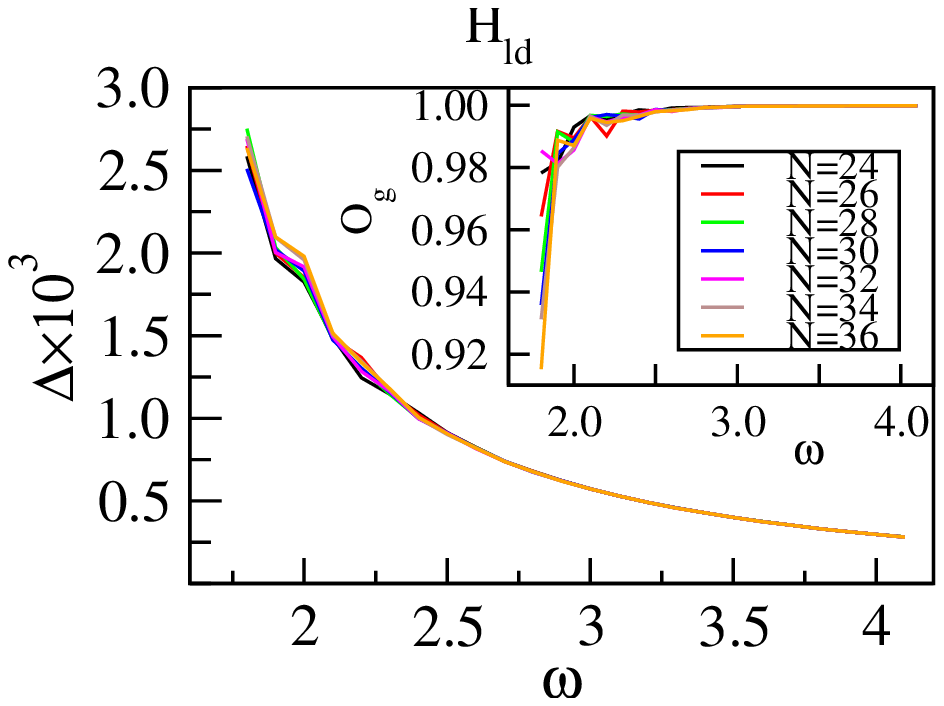}
\includegraphics[height=0.49\columnwidth,angle=270]{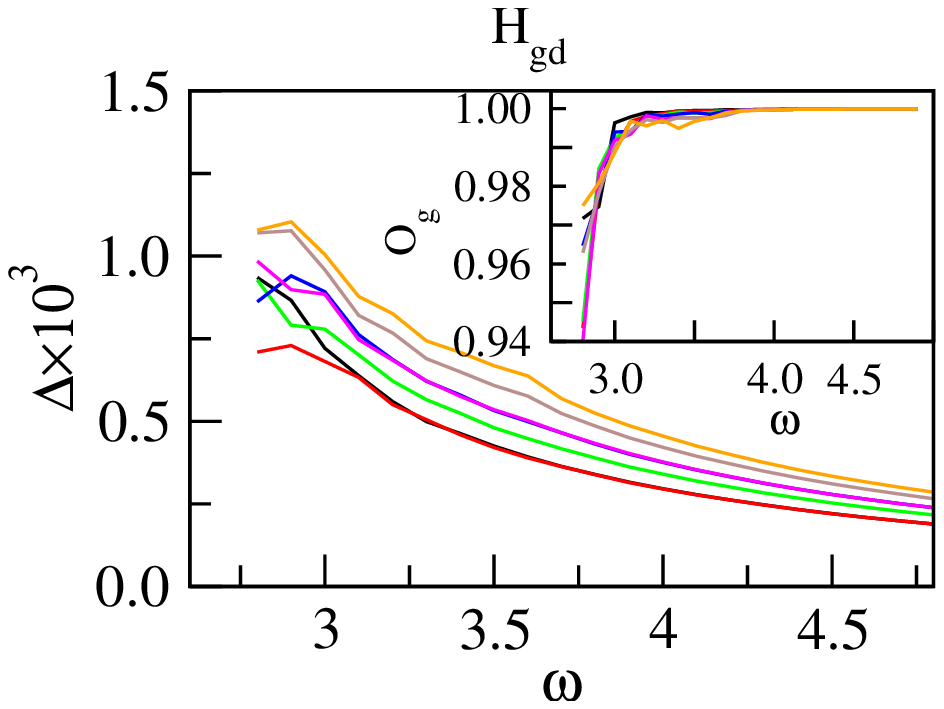}
\caption{Overlap with the ground state $O_g$ and difference of the quasi-energy to the 
ground state energy $\Delta$ as a function of frequency 
for the edge driven and the globally driven model for different $N$.}
\label{overlap}
\end{figure}

In addition to the correlations and number of occupied Floquet modes,
it is also interesting to consider the change of quasi-energies relative 
to the groundstate
 energy $\Delta=\epsilon_q - E_g$ and the magnitude of the wave-function overlap with
the ground state $O_g=|\langle\psi_g|\Phi_0\rangle|$ 
as a function of frequency as shown in Fig.~\ref{overlap}.
The data shows a rather sudden change for $O_g$ as the frequency is lowered, 
fully consistent with the findings in the main text.  

\subsection{Steady state solution: Finding the eigenstate of $\cal H$}
We here adopt a projective method, similar to the standard Rayleigh-Ritz method, 
for solving $K = ({\cal H}-\sigma\bm{1})^2$. 
First we consider a suitable search space 
$\mathcal{V}_m=\{v_1, v_2, \cdots, v_m\}$ of $m$ orthonormal vectors,
 where $m$ is 
much smaller than the actual dimension $d=4M^2(M_1+M_2+1)^2$ of the superblock for the
truncated ${\cal H}$. 
The next step is to project $K$ onto this search space: 
$\tilde{K}_m = V_m^\dag K V_m$, where $V_m$ is the projection matrix whose $i^\text{th}$ column is the vector $v_i$. 
In practice, to form the matrix $\tilde{K}_m$, we first get the new vectors $w_i$'s by applying 
$({\cal H}-\sigma\bm{1})$ to the vectors $v_i$'s: 
$w_i=({\cal H}-\sigma\bm{1})v_i$. Now the $ij^\text{th}$ element of $\tilde{K}_m$ is just the following inner product: 
$(\tilde{K}_m)_{ij}=w_i^\dag w_j$. Any full diagonalization routine can be used to
solve the small projected matrix $\tilde{K}_m$. Let $z$ be one of its eigenvectors. Then, at this stage, the corresponding 
vector, called the Ritz vector,  $V_m z$ is the best approximation of an eigenvector of $K$. Now after forming all the Ritz vectors 
($m$ in number), we pick up the one which has time-averaged maximum overlap with the groundstate. Let $\Phi_{ov}$ be the 
approximate targeted Floquet mode obtained at this stage. The corresponding approximate quasi-energy is given by 
$\epsilon_q=\sigma+\Phi_{ov}^*({\cal H}-\sigma\bm{1})\Phi_{ov}$. 
This approximate eigenstate $\Phi_{ov}$ can now be improved by enlarging the search
space $\mathcal{V}_m$ iteratively until we reach convergence. The enlargement is done by adding
a new linearly independent vector $v_{m+1}$ to form the new search space $\mathcal{V}_{m+1}$ of dimension $m+1$.
A common choice of $v_{m+1}$ is the correction vector, derived from the approximate eigenstate, 
as suggested by Davidson \cite{davidson75}. Let $r$ be the eigenvalue of the projected matrix $\tilde{K}_m$ 
corresponding to the vector $\Phi_{ov}$. Then the correction vector $C$  is defined as 
$C_i = [(K-r)\Phi_{ov}]_i/(r-K_{ii})$, where $K_{ii}$ is the $i^\text{th}$ diagonal element of the matrix $K$. This vector 
$C$ is then orthonormalized with the existing basis vectors of $\mathcal{V}_m$ to form the new basis vector $v_{m+1}$. 
 After this, we construct the new projected matrix $\tilde{K}_{m+1}$ (this time we only need to calculate the 
elements corresponding to the new basis vector). The enlarged matrix $\tilde{K}_{m+1}$ is then diagonalized and the targeted 
mode is obtained from the appropriate eigenvector of the matrix. This iterative process is continued until
the convergence is reached. 
If the dimension of the search space becomes
inconveniently large (say, about 40), we restart the whole process with a few (say, about 10) latest approximate
eigenstates which have largest overlap with the groundstate. During the restart, we also replace the scaling/shifting energy 
$\sigma$ by the approximate quasi-energy found at this stage. We found that this change in scaling energy makes the overall
convergence much faster. 

It may be worth mentioning here that, since we are looking
for a Floquet mode with largest overlap with the groundstate, we begin the diagonalization process with just one vector
$v_1=(\cdots, \psi_g, \cdots)^T$, where the groundstate $\psi_g$ is placed in the block corresponding to $n=0$ Fourier component
while all other elements are taken to be 0.




\subsection{Performance of the proposed Floquet DMRG method}
To verify the performance of the Floquet DMRG method, we first compare the Floquet DMRG results (quasi-energies) with the {\it exact diagonalization}
(ED) results for system sizes N = 20 and 22. For the ED calculations, we solve the truncated ${\cal H}$ directly
using the algorithm described above. For the Floquet DMRG calculations, we start with system size $N=6$ and then grow the
system following the Floquet DMRG algorithm stated in the main text. The results can be seen in the Table \ref{table1} for both $H_{ld}$ and $H_{gd}$,
where $\epsilon_q^{ED}$ ($\epsilon_q$) is the quasi-energy of the Floquet mode with  largest ground-state overlap as found
using ED (DMRG) method. To compare the quasi-energies relative to the corresponding groundstate energies $E_g$, the
latter quantities are also provided in the Table. For representative parameter values, the accuracy in calculating the quasi-energies is
found to be of the order of $10^{-7}$. 

\begin{table}[h!]
\begin{center}
\begin{tabular}{c c c c c}
\hline
~& $E_g$& $\epsilon_q^{ED}$& $|\epsilon_q^{ED}-\epsilon_q|$\\  \hline
$H_{ld}$(N=20)&-8.68247333&~ -8.68189932 &~$1.1\times10^{-7}$ \\
$H_{ld}$(N=22)&-9.56807587&~ -9.56750213 &~$1.2\times10^{-7}$\\
$H_{gd}$(N=20)&-8.68247333&~ -8.68216072 &~$1.1\times10^{-7}$\\
$H_{gd}$(N=22)&-9.56807587&~ -9.56773125 &~$5.2\times10^{-7}$\\ \hline
\end{tabular}
\caption{The calculations are done with $M_1=3$, $M_2=5$, and $m=180$. For $H_{ld}$, we take $\mu = 0.1J$ and $\omega = 3J$, and for $H_{gd}$, 
we take $\mu = 0.02J$ and $\omega = 3.5J$.}
\label{table1}
\end{center}
\end{table}

\begin{figure}[t]
\includegraphics[height=9.5cm,angle=270]{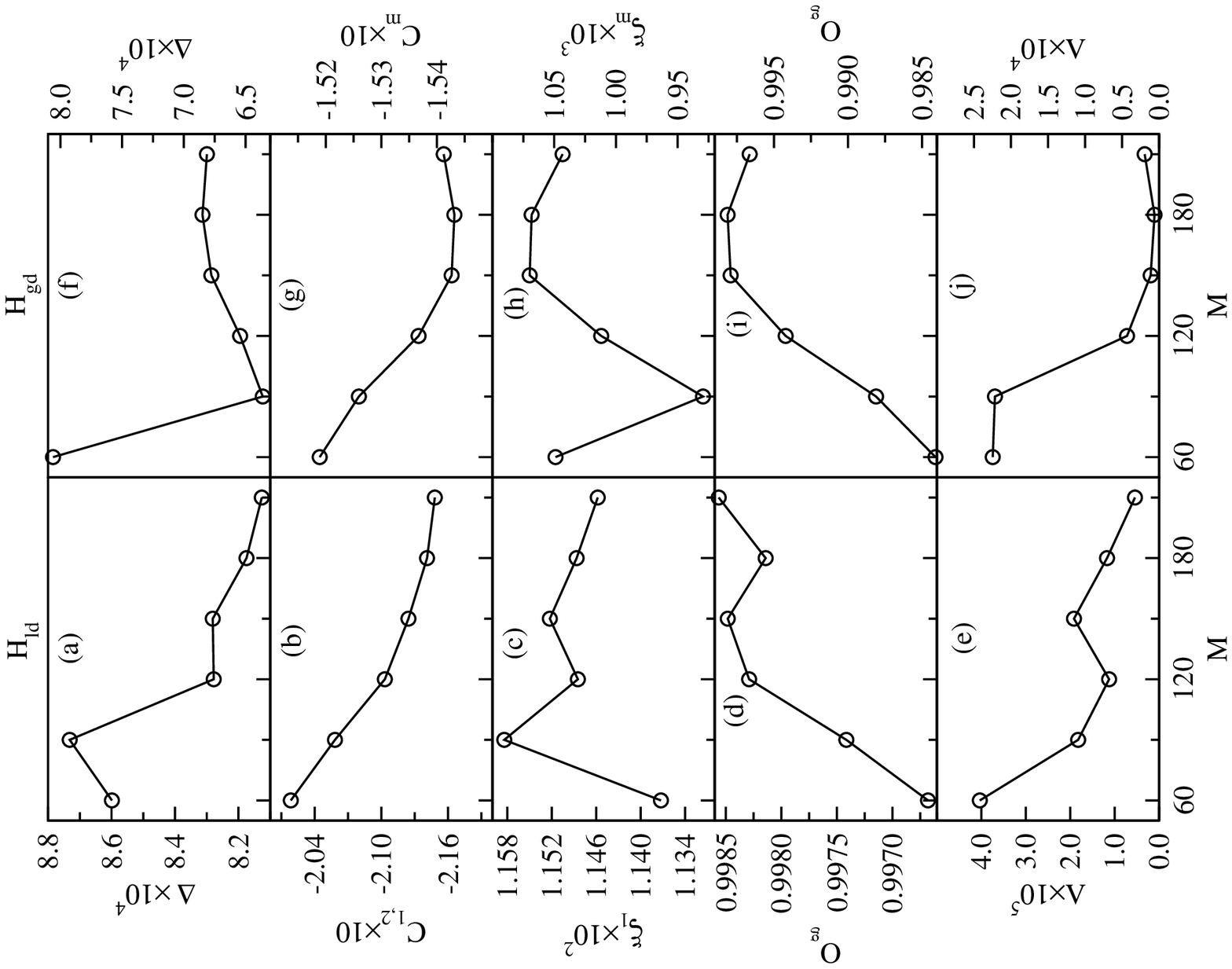}
\caption{Different quantities for $N=32$ are plotted as function of $M$ 
for edge and globally driven systems 
in (a)-(e) and (f)-(j), respectively. We take $\omega=2.6J$ and $\mu=0.1J$ for the 
edge driven system $H_{ld}$, and $\omega=3.2J$ and $\mu=0.02J$ 
for the globally driven system $H_{gd}$.}
\label{cnvrg}
\end{figure}

It may be stressed here that 
finding the Floquet mode  
by solving ${\cal H}$ does not follow the variational principle;
as a consequence, the accuracy of a calculated quantity does not always increase monotonically with $M$ (maximum number of states retained at
each Floquet DMRG step). However, overall results get better with increasing $M$, as can be seen in Fig. \ref{cnvrg}.
For the locally driven system $H_{ld}$, the difference of the quasi-energy from the groundstate energy $\Delta=\epsilon_q - E_g$, the time-averaged  
correlation between the first two sites $C_{1,2}$, 
the temporal fluctuation of the first site $\xi_1$, 
the time-averaged overlap $O_g$ of the 
targeted mode with the groundstate and the DMRG truncation error 
$\Lambda$ are shown respectively in Fig.~\ref{cnvrg} (a) to (e). For the globally driven system $H_{gd}$, we show the same plots in Fig.~\ref{cnvrg} (f)-(j), except that 
this time we plot the average nearest-neighbor correlation ${C}_m$ and the average temporal fluctuation ${\xi}_m$  in the middle (m)      
instead of $C_{1,2}$ and $\xi_1$ respectively.\\
In the following, we estimate the error associated with the DMRG calculations (for $M=180$; since we obtained our main results keeping 
$M=180$). Let $A_{150}$, $A_{180}$ and $A_{210}$ are values of a quantity $A$  when we keep $M$ = 150, 180 and 210 respectively; then the 
average change in $A$ due to change in $M$ by 30 is $\frac12(|A_{180}-A_{150}|+|A_{210}-A_{180}|)$. 
In each DMRG step the total system is enlarged by two spins, so that each block (E and S) 
has size $2M$,
which is then again reduced to the $M$ states with the largest density 
matrix eigenvalues (discarding the other $M$ states).
The relative error for 
discarding $M=180$ states can then be estimated from 
$\frac12(|A_{180}-A_{150}|+|A_{210}-A_{180}|)\times\frac{180}{30}$. 
Accordingly for the locally driven 
system, the errors for the quantities $\epsilon_q$, $C_{1,2}$, $\xi_1$ and $O_g$ are estimated to be 3.32$\times 10^{-4}\%$, 3.33\%, 1.69\% 
and 0.23\% respectively. Similarly for the globally
driven system, the error estimations are 2.34$\times 10^{-4}\%$, 0.47\%, 7.49\% and 0.51\% respectively for the quantities 
$\epsilon_q$, ${C}_m$, ${\xi}_m$ and $O_g$. It may be mentioned here that, to verify the 
performance of the DMRG method, we deliberately chose $\omega$ values in the intermediate/ moderate range (where system goes from the high-frequency  
localized phase to the low-frequency ergodic phase). The accuracy of a calculated quantity gets better as we move towards a high-frequency regime.

\subsection{Derivation of the effective time-independent Hamiltonian}

Let us consider the general periodically driven time-dependent Hamiltonian 
\begin{align} 
H(t)&=J\sum_{j=1}^{N-1} \left[\frac{1}{2}\left(S^+_j S^-_{j+1} +S^-_j S^+_{j+1}\right)+S^z_j S^z_{j+1}\right]+ 2\sum_{j=1}^{N}\mu_j ~\textrm{cos} (\omega t) S^z_{j} \label{hs_gd2}. 
\end{align}
Its steady-state solution is of the form $|\psi(t)\rangle=e^{-i \epsilon t}\,|\Phi(t)\rangle$ where $|\Phi(t)\rangle=|\Phi(t+T)\rangle$ is time-periodic in $T=\frac{2\pi}{\omega}$ and fulfills the eigenvalue equation
\begin{align}\label{floquet_eveq}
\left(H(t)-i  \frac{\partial}{\partial t}\right)|\Phi(t)\rangle=\epsilon |\Phi(t)\rangle.
\end{align}
In solving Eq.~(\ref{floquet_eveq}) in the large $\omega/\mu$ limit we follow the lines of Ref.~\onlinecite{wang14}. We first consider the simplified Hamiltonian 
\begin{align} 
H_0(t)&=J\sum_{j=1}^{N-1} S^z_j\cdot S^z_{j+1}+ 2\sum_{j=1}^{N}\mu_j ~\textrm{cos} (\omega t) S^z_{j} \label{hs_ising}
\end{align}
which is diagonal in the Ising states $|\{{s_j^z}\}\rangle=\prod_{\otimes j} |{s_j^z}\rangle$. Here, $|s_j^z\rangle_j$ represents the local $S^z_j$-basis state, i.e.~$S_j^z |s_j^z\rangle=s_j^z |s_j^z\rangle $. Consequently, 
the corresponding simplified Floquet equation 
\begin{align}
\left(H_0(t)-i  \frac{\partial}{\partial t}\right)|\Phi(t)\rangle=\epsilon |\Phi(t)\rangle
\end{align}\label{floquet_eveq_ising}
can be directly solved yielding solutions
\begin{align}\label{states_ising}
|\Phi^\ell (t)\rangle=e^{i \ell \omega t} \prod_{\otimes j} e^{- i \frac{2 \mu_j}{\omega}s_j^z \sin(\omega t)} |s_j^z\rangle
\end{align} and quasi-energies
\begin{align}
\epsilon_\ell=J\sum_j s_j^z s_{j+1}^z+ \ell  \omega.
\end{align}
Note, that we used the (trivial) 
index $\ell$ to extend the solutions over all values of $\epsilon$
analogous to the extended zone scheme for Bloch waves, which will later allow us to 
calculate matrix elements of solutions from different Floquet bands labeled by $\ell$.
However, the physical relevant steady state solution $|\Phi (t)\rangle
= |\Phi^0(t)\rangle$ corresponds to $\ell=0$ in  Eq.~(\ref{states_ising}). 
 Next, we take the non-diagonal spin-flip terms into account. For the effective Hamiltonian, we are interested in the time-averaged behavior. Therefore, we define the time-averaged scalar product
\begin{align}
\langle\langle u_1(t)|u_2(t)\rangle\rangle=\frac{1}{T} \int_0^{T} dt \langle u_1(t)|u_2(t)\rangle
\end{align} 
and determine the corresponding matrix elements of the full Floquet Hamiltonian
 with respect to any two Floquet states  $\Phi^{\ell_1}_1$ and $\Phi^{\ell_2}_2$ 
defined in Eq.~(\ref{states_ising})
in any Floquet bands, i.e.
\begin{align}
\mathcal{H}_{\Phi_1^{\ell_1}, \Phi^{\ell_2}_2}:=\langle\langle \Phi^{\ell_1}_1 | \left(H(t)-i  \frac{\partial}{\partial t}\right)  |\Phi^{\ell_2}_2 \rangle \rangle.
\end{align}
We obtain in terms of the respective quantum-numbers ($\{s_j^z\}_1$ and $\{s_j^z\}_2$)
\begin{equation}
\mathcal{H}_{\Phi_1^{\ell_1}, \Phi^{\ell_2}_2} =
\langle \langle \Phi_1^{\ell_1}               |\sum_i  \frac{1}{2}\left(S^+_i S^-_{i+1} +S^-_i S^+_{i+1}\right) |\Phi^{\ell_2}_2\rangle \rangle
+ \delta_{\ell_1,\ell_2} 
 \delta_{\{s_j^z\}_1,\{s_j^z\}_2} \epsilon_{\ell_1}. 
\end{equation}
Note that 
the non-diagonal contributions $ \langle\langle \Phi_1^{\ell_1}|S^+_i S^-_{i+1}  |\Phi^{\ell_2}_2\rangle \rangle $
	are only non-zero if $\{s_j^z\}_1=\{s_j^z\}_2$ for $j\neq i,i+1$ and $\{s_i^z\}_1=\{s_i^z\}_2+1$  and $\{s^z_{i+1}\}_1=\{s^z_{i+1}\}_2-1$. The corresponding phase factors are
	\begin{align}
	\frac{1}{T} \int_0^{T} dt \; e^{i (\ell_2-\ell_1)\omega t} e^{i \frac{2 (\mu_i-\mu_{i+1})}{\omega} \sin(\omega t)}=J_{\ell_2-\ell_1}\left(\frac{2 (\mu_i-\mu_{i+1})}{\omega}\right)
	\end{align} 
	where $J_\ell(z)$ denotes the Bessel function of the first kind. Taking into account only terms with $\ell_1=\ell_2$ is justified in the high frequency limit and
gives the effective time-independent Hamiltonian 
	\begin{align}\label{h_effective2} 
	H_{\rm eff}&=\sum_{j=1}^{N-1} \left[\frac{J_{j,j+1}}{2}\left(S^+_j S^-_{j+1} +S^-_j S^+_{j+1}\right)+ J S^z_jS^z_{j+1}\right]  
	\end{align}
	with $J_{j,j+1}=J J_0\left(\frac{2 (\mu_j-\mu_{j+1})}{\omega}\right)$.  
\end{document}